\documentclass[11pt]{amsart}

\numberwithin{equation}{section}
\usepackage{hyperref}
\begin{document}

\title{Electrodynamic forces in elastic matter}
\thanks{Nuovo Cimento, in press.}
\author{S. Antoci \and L. Mihich}
\address{Dipartimento di Fisica ``A. Volta'' and I. N. F. M.,
    Via Bassi 6, Pavia, Italy}
\email{Antoci@fisav.unipv.it}
\keywords{Classical field theory. General relativity. Electromagnetism}

\begin{abstract}
A macroscopic theory for the dynamics of elastic, isotropic matter
in presence of electromagnetic fields is proposed here. We avail
of Gordon's general relativistic derivation of Abraham's
electromagnetic energy tensor as starting point. The necessary
description of the elastic and of the inertial behaviour of matter
is provided through a four-dimensional generalisation of Hooke's
law, made possible by the introduction of a four-dimensional
``displacement'' vector. As intimated by Nordstr\"om, the physical
origin of electrostriction and of magnetostriction is attributed
to the change in the constitutive equation of electromagnetism
caused by the deformation of matter. The part of the
electromagnetic Lagrangian that depends on that deformation is
given explicitly for the case of an isotropic medium and the
resulting expression of the electrostrictive force is derived,
thus showing how more realistic equations of motion for matter
subjected to electromagnetic fields can be constructed.
\end{abstract}

%%% ----------------------------------------------------------------------
\maketitle
%%% ----------------------------------------------------------------------
\section{Introduction}
\label{I} According to a widespread belief cultivated by
present-day physicists, general relativity exerts its sovereign
power in the heavens, where it supposedly rules tremendous
astrophysical processes and awesome cosmological scenarios, but it
has essentially nothing to say about more down to earth issues
like the physics of ordinary matter, as it shows up in terrestrial
laboratories. This way of thinking does not conform to the hopes
expressed by Bernhard Riemann in his celebrated {\it
Habilitationsschrift} \cite{Riemann1868}. While commenting upon
the possible applications to the physical space of his new
geometrical ideas, he wrote:
\begin{quotation}
\noindent Die Fragen \"uber das Unme{\ss}bargro{\ss}e sind f\"ur
die Naturerkl\"arung m\"usige Fragen. Anders verh\"alt es sich
aber mit den Fragen \"uber das Unme{\ss}bar\-kleine \footnote{The
questions about the infinitely great are for the interpretation of
nature useless questions. But this is not the case with the
questions about the infinitely small.}.
\end{quotation}
The latter is presently supposed to be the exclusive hunting
ground for quantum physics, whose workings occur at their best in
the flat space of Newton. The classical field theories, in
particular classical electromagnetism, are believed to have
accomplished their midwife task a long time ago. Although they are
still revered, since they act as cornerstones for the applied
sciences, and also provide the very foundation on which quantum
mechanics and quantum field theory do stand, no fundamental
insights are generally expected from their further frequentation.
This mind habit cannot subsist without a strenuous act of faith in
the final nature of the present-day reductionist programme: since
for all practical purposes we have eventually attained the right
microscopic laws, getting from there the right macroscopic physics
should be just a matter of deduction by calculation (for the ever
growing army of computer addicts, a sheer problem of computing
power). Given time and endurance, we should be able to account for
all the observed phenomena just by starting from our very simple
microscopic laws!\newline\indent It is not here the place for
deciding how much this bold faith in the capabilities of today's
reductionist approach be strenghtened by its undeniable successes,
and how much it depend on having tackled just the sort of problems
that are most suited to such a method. However, when confronted
with the end results of many reductionist efforts, the obdurate
classicist cannot help frowning in puzzlement. While he expects to
meet with macroscopic laws derived from the underlying microscopic
postulates by a pure exercise of logic, the everyday's practice
confronts him with a much less palatable food. At best he is
presented with rather particular examples usually worked out from
the sacred principles through the surreptitious addition of a host
of subsidiary assumptions. In the worst cases he is forced to
contemplate and believe sequels of colourful plots and diagrams,
generated by some computing device in some arcane way that he is
simply impotent at producing again. In the intention of their
proponents, both the ``analytic'' instances and their numerical
surrogates should provide typical examples of some supposedly
general behaviour, really stemming from the basic tenets of the
theory, and in many a case this lucky occurrence may well have
occurred, since ``God watches over applied mathematicians''
\cite{Truesdell87}. Nevertheless, the longing of the classicist
for macroscopic laws of clear conceptual ancestry that do
encompass in surveyable form a large class of phenomena remains
sadly disattended. He is led to remind of the pre-quantum era,
when both the reductionist approaches and the macroscopic ones
were believed to be equally important tools for the advancement of
physical knowledge \footnote{It is remarkable how the otherwise
daring Min\-kow\-ski kept a cautious attitude with respect to
Lorentz' atomistic theory of electricity both in his fundamental
paper \cite{Minkowski08} of 1908 and also in his ``Nachlass''
notes \cite{Minkowski10}, posthumously edited and published by Max
Born.}, and to wonder whether relegating the macroscopic field
approach to the ``phenomenological'' dustbin was really a wise
move. Before being removed from center stage by more modern
approaches, the macroscopic field theory lived long enough for
developing, in the hands of great natural philosophers and
mathematicians, theoretical tools of a very wide scope that, if
still remembered and cultivated, would be recognized to be very
useful today.

\section{Electrodynamic forces in material media}
\label{sec:2} One of the clearest instances in which recourse to
macroscopic field theory proves to be a quite helpful research
tool occurs when one tries to describe the electromagnetic forces
in material media. Since the time of Lorentz this has been a very
challenging task; reductionist approaches starting from classical
mechanics and from vacuum electrodynamics, for reasons clearly
spelled out {\it e. g.} by Ott \cite{Ott52}, end up in a
disappointing gamut of possibilities also when the program of a
rigorous special relativistic derivation is tenaciously adhered to
\cite{DeGroot67}, \cite{DeGroot68}. To our knowledge, a derivation
of the macroscopic forces exerted by the electromagnetic field on
a material medium performed by availing of quantum electrodynamics
as the  underlying microscopic theory, that should be {\it de
rigueur} in the reductionist programme, has never been undertaken
\footnote{By availing once more of the midwife abilities of
classical field theory, the converse has instead been attempted:
some forms of ``phenomenological'' classical electrodynamics in
matter has been subjected to some quantisation process
\cite{Jauch48}, \cite{Nagy55}, \cite{BL70}, thereby producing
diverse brands of ``phenomenological'' photons.}.\par Happily
enough, the theoretical advance in the methods for producing the
stress energy momentum tensor of non gravitational fields occurred
with the onset of general relativity theory \cite{Hilbert15},
\cite{Klein17} have allowed W. Gordon to find, through a clever
reduction to the vacuum case of the latter theory \cite{Gordon23},
a clear-cut argument for determining the electromagnetic forces in
matter that is homogeneous and isotropic in its local rest frame.
We shall recall Gordon's result in the next Section, since
extending his outcome to the case of an elastic medium is just the
scope of the present paper.

\section{Gordon's {\it reductio ad vacuum} of the constitutive
equation of electromagnetism} \label{sec:3} We adopt hereafter
Gordon's conventions \cite{Gordon23} and assume that the metric
tensor $g_{ik}$ can be locally brought to the diagonal form
\begin{equation}\label{3.1}
g_{ik}=\eta_{ik}\equiv diag(1,1,1,-1)
\end{equation}
at a given event through the appropriate transformation of
coordinates. According to the established convention \cite{Post62}
let the electric displacement and the magnetic field be
represented by the antisymmetric, contravariant tensor density
${\bf H}^{ik}$, while the electric field and the magnetic
induction are accounted for by the skew, covariant tensor
$F_{ik}$. With these geometrical objects we define the
four-vectors:
\begin{equation}\label{3.2}
F_{i}=F_{ik}u^{k},~~~H_{i}=H_{ik}u^{k},
\end{equation}
where $u^i$ is the four-velocity of matter. In general relativity
a linear electromagnetic medium can be told to be homogeneous and
isotropic in its rest frame if its constitutive equations can be
written as
\begin{equation}\label{3.3}
\mu{H^{ik}}=F^{ik}+(\epsilon\mu-1)(u^{i}F^{k}-u^{k}F^{i}),
\end{equation}
where the numbers $\epsilon$ and $\mu$ account for the dielectric
constant and for the magnetic permeability of the medium
\cite{Gordon23}. This equation provides the constitutive relation
in the standard form:
\begin{equation}\label{3.4}
{\bf H}^{ik}={\frac{1}{2}}{\bf X}^{ikmn}F_{mn},
\end{equation}
valid for linear media \cite{Nordstroem23}. Gordon observed that
equation (\ref{3.3}) can be rewritten as
\begin{equation}\label{3.5}
\mu{H^{ik}}=\big[g^{ir}-(\epsilon\mu-1)u^{i}u^{r}\big]
\big[g^{ks}-(\epsilon\mu-1)u^{k}u^{s}\big]F_{rs}.
\end{equation}
By introducing the ``effective metric tensor''
\begin{equation}\label{3.6}
\sigma^{ik}=g^{ik}-(\epsilon\mu-1)u^{i}u^{k},
\end{equation}
the constitutive equation takes the form
\begin{equation}\label{3.7}
\mu{\bf H}^{ik}=\sqrt{g}\sigma^{ir}\sigma^{ks}F_{rs},
\end{equation}
where $g\equiv-\det(g_{ik})$. The inverse of $\sigma^{ik}$ is
\begin{equation}\label{3.8}
\sigma_{ik}=g_{ik}+\big(1-{\frac{1}{\epsilon\mu}}\big)u_{i}u_{k},
\end{equation}
and one easily finds \cite{Gordon23} that
\begin{equation}\label{3.9}
\sigma=\frac{g}{\epsilon\mu},
\end{equation}
where $\sigma\equiv-\det(\sigma_{ik})$. Therefore the constitutive
equation can be eventually written as
\begin{equation}\label{3.10}
{\bf H}^{ik}=\sqrt{\frac{\epsilon}{\mu}}
\sqrt{\sigma}\sigma^{ir}\sigma^{ks}F_{rs}.
\end{equation}
\section{Gordon's derivation of Abraham's energy tensor}
\label{sec:4} This result is the basis of Gordon's argument:
since, apart from the constant factor $\sqrt{\epsilon/\mu}$,
equation (\ref{3.10}) is the constitutive equation for
electromagnetism in a general relativistic vacuum whose metric be
$\sigma_{ik}$, the Lagrangian density that we shall use for
deriving the laws of the field is:
\begin{equation}\label{4.1}
{\bf L}=\frac{1}{4}\sqrt{\frac{\epsilon}{\mu}}
\sqrt{\sigma}F^{(i)(k)}F_{ik}-{\bf s}^{i}\varphi_{i},
\end{equation}
where ${\bf s}^{i}$ is the four-current density, while
$\varphi_{i}$ is the potential four-vector that defines the
electric field and the magnetic induction:
\begin{equation}\label{4.2}
F_{ik}\equiv \varphi_{k,i}-\varphi_{i,k}.
\end{equation}
We have adopted the convention of enclosing within round brackets
the indices that are either moved with $\sigma_{ik}$ and
$\sigma^{ik}$, or generated by performing the Hamiltonian
derivative with respect to the mentioned tensors. The position
(\ref{4.2}) is equivalent to asking the satisfaction of the
homogeneous set of Maxwell's equations
\begin{equation}\label{4.3}
F_{[ik,m]}=0,
\end{equation}
while equating to zero the variation of the action integral
$\int{\bf L}d\Omega$ with respect to $\varphi_i$ entails the
fulfilment of the inhomogeneous Maxwell's set
\begin{equation}\label{4.4}
{\bf H}^{ik}_{~~,k}={\bf s}^{i}.
\end{equation}
In our general relativistic framework, we can avail of the results
found by Hilbert and Klein \cite{Hilbert15}, \cite{Klein17} for
determining the energy tensor of the electromagnetic field. If the
metric tensor of our pseudo-Riemannian space-time were
$\sigma_{ik}$, Hilbert's method would provide the electromagnetic
energy tensor by executing the Hamiltonian derivative of the
Lagrangian density ${\bf L}$ with respect to that metric:
\begin{equation}\label{4.5}
\delta{\bf L}\equiv{\frac{1}{2}} {\bf
T}_{(i)(k)}\delta\sigma^{ik},
\end{equation}
and we would get the mixed tensor density
\begin{equation}\label{4.6}
{\bf T}_{(i)}^{~~~(k)}=F_{ir}{\bf H}^{kr}
-{\frac{1}{4}}\delta_i^{~k}F_{rs}{\bf H}^{rs},
\end{equation}
which is just the general relativistic form of the energy tensor
density proposed by Minkowski in his fundamental work
\cite{Minkowski08}. But $g_{ik}$, not $\sigma_{ik}$, is the true
metric that accounts for the structure of space-time and, through
Einstein's equations, defines its overall energy tensor. Therefore
the partial contribution to that energy tensor coming from the
electromagnetic field must be obtained by calculating the
Hamiltonian derivative of ${\bf L}$ with respect to $g_{ik}$.
After some algebra \cite{Gordon23} one easily gets the
electromagnetic energy tensor:
\begin{equation}\label{4.7}
T_{i}^{~k}=F_{ir}H^{kr}-{\frac{1}{4}}
\delta_{i}^{~k}F_{rs}H^{rs}-(\epsilon\mu-1)\Omega_{i}u^{k},
\end{equation}
where
\begin{equation}\label{4.8}
\Omega^{i}\equiv-\big(T_{k}^{~i}u^{k}+u^{i}T_{mn}u^{m}u^{n}\big)
\end{equation}
is Minkowski's ``Ruh-Strahl'' \cite{Minkowski08}. Since
$\Omega^{i}u_{i}\equiv0$, substituting (\ref{4.7}) into
(\ref{4.8}) yields:
\begin{equation}\label{4.9}
\Omega^{i}=F_{m}H^{im}-F_{m}H^{m}u^{i}
=u_{k}F_{m}\big(H^{ik}u^{m}+H^{km}u^{i}+H^{mi}u^{k}\big),
\end{equation}
and one eventually recognizes that $T_{ik}$ is the general
relativistic extension of Abraham's tensor \cite{Abraham09} for a
medium that is homogeneous and isotropic when looked at in its
local rest frame. The four-force density exerted by the
electromagnetic field on the medium shall be given by (minus) the
covariant divergence of that energy tensor density:
\begin{equation}\label{4.10}
{\bf f}_{i}=-{\bf T}^{~k}_{i~;k},
\end{equation}
where the semicolon stands for the covariant differentiation done
by using the Christoffel symbols built with the metric $g_{ik}$.
Abraham's energy tensor is an impressive theoretical outcome, that
could hardly have been anticipated on the basis of heuristic
arguments. Quite remarkably, the so called Abraham's force
density, that stems from the four-divergence of that tensor, has
found experimental confirmation in some delicate experiences
performed by G. B. Walker et al. \cite{Walker75a},
\cite{Walker75b}. Despite this, Abraham's rendering of the
electrodynamic forces is not realistic enough, for it does not
cope with the long known phenomena of electrostriction and of
magnetostriction. We need to find its generalization, and we shall
start from considering the case of linear elastic media, to which
Hooke's law applies. This task would be made formally easier if
one could avail of a relativistic reformulation of the linear
theory of elasticity; the next Section will achieve this goal
through a four-dimensional formulation of Hooke's law
\cite{Antoci99} that happens to be rather well suited to our
scopes.
\section{A four-dimensional formulation of Hooke's law}
\label{sec:5} By availing of Cartesian coordinates and of the
three-dimensional tensor formalism, that was just invented to cope
with its far-reaching consequences, Hooke's law {\it ``ut tensio
sic vis''} \cite{Hooke1678} can be written as
\begin{equation}\label{5.1}
{\Theta}^{\lambda\mu}={\frac{1}{2}}C^{\lambda\mu\rho\sigma}
(\xi_{\rho,\sigma}+\xi_{\sigma,\rho}),
\end{equation}
where ${\Theta}^{\lambda\mu}$ is the three-dimensional tensor that
defines the stresses arising in matter due to its displacement,
given by the three-vector $\xi^\rho$, from a supposedly relaxed
condition, and $C^{\lambda\mu\rho\sigma}$ is the constitutive
tensor whose build depends on the material features and on the
symmetry properties of the elastic medium. It seems natural to
wonder whether this venerable formula can admit of not merely a
redressing, but of a true generalization to the four-dimensions of
the general relativistic spacetime. From a formal standpoint, the
extension is obvious: one introduces a four-vector field $\xi^i$,
that should represent a four-dimensional ``displacement'', and
builds the ``deformation'' tensor
\begin{equation}\label{5.2}
S_{ik}={1\over2}(\xi_{i;k}+\xi_{k;i}).
\end{equation}
A four-dimensional ``stiffness'' tensor density ${\bf{C}}^{iklm}$
is then introduced; it will be symmetric in both the first pair
and the second pair of indices, since it will be used for
producing a ``stress-momentum-energy'' tensor density
\begin{equation}\label{5.3}
{\bf{T}}^{ik}={\bf{C}}^{iklm}S_{lm},
\end{equation}
through the four-dimensional generalization of equation
(\ref{5.1}). It has been found \cite{Antoci99} that this
generalization can be physically meaningful, since it allows one
to encompass both inertia and elasticity in a sort of
four-dimensional elasticity. Let us consider a coordinate system
such that, at a given event, equation (\ref{3.1}) holds, while the
Christoffel symbols are vanishing and the components of the
four-velocity of matter are:
\begin{equation}\label{5.4}
u^1=u^2=u^3=0,\ u^4=1.
\end{equation}
We imagine that in such a coordinate system we are able to
measure, at the chosen event, the three components of the
(supposedly small) spatial displacement of matter from its relaxed
condition, and that we adopt these three numbers as the values
taken by $\xi^\rho$ in that coordinate system, while the reading
of some clock ticking the proper time and travelling with the
medium will provide the value of the ``temporal displacement''
$\xi^4$ in the same coordinate system. By applying this procedure
to all the events of the manifold where matter is present and by
reducing the collected data to a common, arbitrary coordinate
system, we can define the vector field $\xi^i(x^k)$. From such a
field we shall require that, when matter is not subjected to
ordinary strain and is looked at in a local rest frame belonging
to the ones defined above, it will exhibit a ``deformation
tensor'' $S_{ik}$ such that its only nonzero component will be
$S_{44}=\xi_{4,4}=-1$. This requirement is met if we define the
four-velocity of matter through the equation
\begin{equation}\label{5.5}
\xi^i_{;k}u^k=u^i.
\end{equation}
The latter definition holds provided that
\begin{equation}\label{5.6}
det(\xi^i_{;k}-\delta^i_k)=0,
\end{equation}
and this shall be one equation that the field $\xi^i$ must
satisfy; in this way the number of independent components of
$\xi^i$ will be reduced to three \footnote{The fulfilment of
equation (\ref{5.5}) is only a necessary, not a sufficient
condition for the field $\xi^i$ to take up the tentative meaning
that was envisaged above. The physical interpretation of the field
$\xi^i$ can only be assessed {\it a posteriori} from the solutions
of the field equations.}. A four-dimensional ``stiffness'' tensor
$C^{iklm}$ possibly endowed with physical meaning can be built as
follows. We assume that in a locally Minkowskian rest frame the
only nonvanishing components of $C^{iklm}$ are:
$C^{\lambda\nu\sigma\tau}$, with the meaning of ordinary elastic
moduli, and
\begin{equation}\label{5.7}
C^{4444}=-\rho,
\end{equation}
where $\rho$ measures the rest density of matter. But of course we
need defining the four-dimensional ``stiffness'' tensor in an
arbitrary co-ordinate system. The task can be easily accomplished
if the unstrained matter is isotropic when looked at in a locally
Minkowskian rest frame, and this is just the occurrence that we
have already studied from the electromagnetic standpoint in
Sections \ref{sec:3} and \ref{sec:4}. Let us define the auxiliary
metric
\begin{equation}\label{5.8}
\gamma^{ik}=g^{ik}+u^iu^k;
\end{equation}
then the part of $C^{iklm}$ accounting for the ordinary elasticity
of the isotropic medium can be written as \cite{Choquet73}
\begin{equation}\label{5.9}
C^{iklm}_{el.}=-\lambda\gamma^{ik}\gamma^{lm}
-\mu(\gamma^{il}\gamma^{km}+\gamma^{im}\gamma^{kl}),
\end{equation}
where $\lambda$ and $\mu$ are assumed to be constants. The part of
$C^{iklm}$ that accounts for the inertia of matter shall read
instead
\begin{equation}\label{5.10}
C^{iklm}_{in.}=-\rho u^iu^ku^lu^m.
\end{equation}
The elastic part $T^{ik}_{el.}$ of the energy tensor is orthogonal
to the four-velocity, as it should be \cite{Cattaneo71}; thanks to
equation (\ref{5.5}) it reduces to
\begin{eqnarray}\label{5.11}
T^{ik}_{el.}=C^{iklm}_{el.}S_{lm}
=-\lambda(g^{ik}+u^iu^k)(\xi^m_{;m}-1)\nonumber\\
-\mu[\xi^{i;k}+\xi^{k;i}+u_l(u^i\xi^{l;k}+u^k\xi^{l;i})],
\end{eqnarray}
while, again thanks to equation (\ref{5.5}), the inertial part of
the energy tensor proves to be effectively so, since
\begin{equation}\label{5.12}
T^{ik}_{in.}=C^{iklm}_{in.}S_{lm} =\rho u^iu^k.
\end{equation}
The energy tensor defined by summing the contributions
(\ref{5.11}) and (\ref{5.12}) encompasses both the inertial and
the elastic energy tensor of an isotropic medium; when the
macroscopic electromagnetic field is vanishing it should represent
the overall energy tensor, whose covariant divergence must vanish
according to Einstein's equations \cite{Hilbert15},
\cite{Klein17}:
\begin{equation}\label{5.13}
T^{ik}_{;k}=0.
\end{equation}
Imposing the latter condition allows one to write the equations of
motion for isotropic matter subjected to elastic strain
\cite{Cattaneo71}. We show this outcome in the limiting case when
the metric is everywhere flat and the four-velocity of matter is
such that $u^\rho$ can be dealt with as a first order
infinitesimal quantity, while $u^4$ differs from unity at most for
a second order infinitesimal quantity. Also the spatial components
$\xi^\rho$ of the displacement vector and their derivatives are
supposed to be infinitesimal at first order. An easy calculation
\cite{Antoci99} then shows that equation (\ref{5.6}) is satisfied
to the required first order, and that equations (\ref{5.13})
reduce to the three equations of motion:
\begin{equation}\label{5.14}
\rho\xi^\nu_{,4,4}=\lambda\xi^{\rho,\nu}_{~,\rho}
+\mu(\xi^{\nu,\rho}+\xi^{\rho,\nu})_{,\rho},
\end{equation}
and to the conservation equation
\begin{equation}\label{5.15}
\{\rho u^4u^k\}_{,k}=0,
\end{equation}
{\it i. e.}, to the required order, they come to coincide with the
well known equations of the classical theory of elasticity for an
isotropic medium.
\section{Electrostriction and magnetostriction in isotropic matter}
\label{sec:6} Having provided that portion of the equations of
motion of matter that stems from the inertial and from the elastic
part of the energy tensor, we can go back to the other side of our
problem: finding to what changes the electrodynamic forces
predicted in isotropic matter by Gordon's theory must be subjected
in order to account for electrostriction and for magnetostriction.
Driven by a suggestion found in the quoted paper by Nordstr\"om
\cite{Nordstroem23}, we attribute the physical origin of the
electrostrictive and of the magnetostrictive forces to the changes
that the constitutive relation (\ref{3.4}) undergoes when matter
is strained in some way. If one desires to represent explicitly
the effect of a small spatial deformation on the constitutive
relation of electromagnetism, one can replace (\ref{3.4}) with a
new equation, written in terms of the new tensor density ${\bf
Y}^{ikpqmn}$, that can be chosen to be antisymmetric with respect
to the first pair and to the last pair of indices, symmetric with
respect to the second pair. This tensor density allows one to
rewrite the constitutive relation as follows:
\begin{equation}\label{6.1}
{\bf H}^{ik}={1\over2}{\bf Y}^{ikpqmn}S_{pq}F_{mn},
\end{equation}
where $F_{ik}$ is defined by (\ref{4.2}) and $S_{ik}$ is given by
(\ref{5.2}). For the intended application to isotropic matter it
is convenient to split the equation written above in two terms,
one concerning the unstrained medium, that has already been
examined in Sections \ref{sec:3} and \ref{sec:4}, and one dealing
with the spatial deformation proper. Due to equation (\ref{5.5})
one finds
\begin{equation}\label{6.2}
{1\over 2}u^pu^q(\xi_{p;q}+\xi_{q;p})=u_pu^q\xi^p_{;q}=u_pu^p=-1,
\end{equation}
and the part (\ref{3.5}) of the constitutive equation valid for
the isotropic unstrained medium can be rewritten as:
\begin{equation}\label{6.3}
{\bf H}^{ik}_{(u.)}=-{\sqrt{g}\over\mu}
\big[g^{im}-(\epsilon\mu-1)u^iu^m\big]
\big[g^{kn}-(\epsilon\mu-1)u^ku^n\big] u^pu^qS_{pq}F_{mn}.
\end{equation}
For producing the part of the constitutive equation that deals
with the effects of the spatial deformation proper, we recall that
an arbitrary deformation will bring the medium, which is now
supposed to be isotropic when at rest and unstrained, into a
generic anisotropic condition. When the magnetoelectric effect is
disregarded \footnote{Such an effect indeed exists
\cite{Astrov61}, but it is sufficiently {\it rara avis} to be
neglected in the present context.}, the electromagnetic properties
of an anisotropic medium can be summarised, as shown {\it e. g.}
by Sch\"opf \cite{Schoepf64}, by assigning two symmetric
four-tensors $\zeta_{ik}=\zeta_{ki}$ and
$\kappa_{ik}=\kappa_{ki}$, whose fourth row and column vanish in a
coordinate system in which matter happens to be locally at rest.
This property finds tensorial expression in the equations
\begin{equation}\label{6.4}
\zeta_{ik}u^k=0,~~~\kappa_{ik}u^k=0;
\end{equation}
$\zeta_{ik}$ has the r\^ole of dielectric tensor, while
$\kappa_{ik}$ acts as inverse magnetic permeability tensor. Let
$\eta^{iklm}$ be the Ricci-Levi Civita symbol in contravariant
form, while $\eta_{iklm}$ is its covariant counterpart. Then the
generally covariant expression of the constitutive equation for
the anisotropic medium reads \cite{Schoepf64}:
\begin{equation}\label{6.5}
{\bf H}^{ik}=\sqrt{g}\big[(u^i\zeta^{km}-u^k\zeta^{im})u^n
-{1\over2}\eta^{ikrs}u_r\kappa_{sc}u_d\eta^{cdmn}\big]F_{mn}.
\end{equation}
We shall avail of this equation to account for ${\bf
H}^{ik}_{(s.)}$, i. e. for the part of ${\bf H}^{ik}$ produced,
for a given $F_{mn}$, by the presence of ordinary strain in the
otherwise isotropic medium. The tensors $\zeta^{ik}$ and
$\kappa^{ik}$ will now be given a new meaning: they represent
henceforth only the changes in the dielectric properties and in
the inverse magnetic permeability produced by the presence of
strain. If the medium, as supposed, is thought to be isotropic
when at rest and in the unstrained state, the dependence of
$\zeta^{ik}$ and of $\kappa^{ik}$ on $S_{pq}$ will mimic the
dependence on the four-dimensional deformation tensor exhibited by
the elastic stress in an isotropic medium. One shall in fact
write:
\begin{equation}\label{6.6}
\zeta^{km}=\big[\alpha_1\gamma^{km}\gamma^{pq}
+\alpha_2(\gamma^{kp}\gamma^{mq}
+\gamma^{kq}\gamma^{mp})\big]S_{pq},
\end{equation}
where the constants $\alpha_1$ and $\alpha_2$ specify the
electrostrictive behaviour of the isotropic medium. In the same
way one is led to pose:
\begin{equation}\label{6.7}
\kappa^{km}=\big[\beta_1\gamma^{km}\gamma^{pq}
+\beta_2(\gamma^{kp}\gamma^{mq}
+\gamma^{kq}\gamma^{mp})\big]S_{pq}
\end{equation}
to account for the magnetostrictive behaviour; $\beta_1$ and
$\beta_2$ are again the appropriate magnetostrictive constants for
the isotropic medium. By availing of the definitions (\ref{6.6})
and (\ref{6.7}) one eventually writes
\begin{equation}\label{6.8}
{\bf H}^{ik}_{(s.)}=\sqrt{g}\big[(u^i\zeta^{km}-u^k\zeta^{im})u^n
-{1\over2}\eta^{ikrs}u_r\kappa_{sc}u_d\eta^{cdmn}\big]F_{mn}
\end{equation}
for the part of ${\bf H}^{ik}$ due to the ordinary strain. The
overall ${\bf H}^{ik}$ is:
\begin{equation}\label{6.9}
{\bf H}^{ik}={\bf H}^{ik}_{(u.)}+{\bf H}^{ik}_{(s.)},
\end{equation}
and the two addenda at the right-hand side of this equation are
the right-hand sides of equations (\ref{6.3}) and (\ref{6.8});
therefore the overall constitutive relation has just the form
intimated by equation (\ref{6.1}) for a general medium. As we did
when electrostriction and magnetostriction were neglected, we
assume again that the Lagrangian density ${\bf L}$ for the
electromagnetic field in presence of the four-current density
${\bf s}^i$ shall read:
\begin{equation}\label{6.10}
{\bf L}={1\over 4}{\bf H}^{ik}F_{ik} -{\bf s}^i\varphi_i,
\end{equation}
where $\varphi_i$ is the four-vector potential, while ${\bf
H}^{ik}$ has the new definition (\ref{6.9}). Maxwell's equations
(\ref{4.3}) and (\ref{4.4}) are then obtained in just the same way
as it occurred with the Lagrangian density (\ref{4.1}). Like ${\bf
H}^{ik}$, also ${\bf L}$ can be split into an ``unstrained'' part
${\bf L}_{(u.)}$, given by equation (\ref{4.1}), and a term
stemming from strain, that will be called ${\bf L}_{(s.)}$. The
Hamiltonian differentiation of ${\bf L}_{(u.)}$ with respect to
the metric $g_{ik}$ produces the general relativistic version of
Abraham's energy tensor, as we know from Section \ref{sec:4}. For
clearness, we will rewrite it here as
\begin{equation}\label{6.11}
\big({\bf T}^{ik}\big)_{(u.)}=\sqrt{g}\big[F^i_{~r}H^{kr}_{(u.)}
-{1\over4}g^{ik}F_{rs}H^{rs}_{(u.)}
-(\epsilon\mu-1)\Omega^iu^{k}\big],
\end{equation}
where $\Omega^i$ now reads:
\begin{equation}\label{6.12}
\Omega^{i}=u_{k}F_{m}\big(H^{ik}_{(u.)}u^{m}+H^{km}_{(u.)}u^{i}
+H^{mi}_{(u.)}u^{k}\big).
\end{equation}
Let us now deal with the explicit form of ${\bf L}_{(s.)}$. For
simplicity we shall do so when only electrostriction is present,
{\it i. e.} when $\kappa^{ik}=0$. In this case one writes:
\begin{equation}\label{6.13}
{\bf L}_{(s.)}={1\over 4}{\bf H}^{ik}_{(s.)}F_{ik}
={1\over4}\sqrt{g}\big[(u^i\zeta^{km}-u^k\zeta^{im})u^n
\big]F_{mn}F_{ik},
\end{equation}
where $\zeta^{km}$ is defined by (\ref{6.6}). Due to the
antisymmetry of $F_{ik}$, ${\bf L}_{(s.)}$ can be rewritten as
\begin{eqnarray}\label{6.14}
{\bf L}_{(s.)}={1\over2}\sqrt{g}u^iu^n\zeta^{km}F_{mn}F_{ik}
={1\over2}\sqrt{g}u^iu^n\big[\alpha_1\gamma^{km}\gamma^{pq}\nonumber\\
+\alpha_2(\gamma^{kp}\gamma^{mq}+\gamma^{kq}\gamma^{mp})
\big]S_{pq}F_{mn}F_{ik}.
\end{eqnarray}
Thanks to equation (\ref{5.5}) one finds from (\ref{6.6}):
\begin{equation}\label{6.15}
\zeta^{km} =\alpha_1(g^{km}+u^ku^m)(\xi^s_{;s}-1)
+\alpha_2\big[\xi^{k;m}+\xi^{m;k}
+u_s(u^k\xi^{s;m}+u^m\xi^{s;k})\big],
\end{equation}
hence:
\begin{eqnarray}\label{6.16}
{\bf L}_{(s.)}=-{1\over2}\sqrt{g}\zeta^{km}F_kF_m
={1\over2}\sqrt{g}u^iu^n\
\bigg\{\alpha_1(g^{km}+u^ku^m)(\xi^s_{;s}-1)\nonumber\\
+2\alpha_2\big[\xi^{k;m}+u_su^k\xi^{s;m}\big]\bigg\} F_{ik}F_{mn}.
\end{eqnarray}
But of course the expression $u^iu^kF_{ik}$ identically vanishes;
therefore the previous equation reduces to:
\begin{equation}\label{6.17}
{\bf L}_{(s.)}
={1\over2}\sqrt{g}u^au^n\big[\alpha_1g^{bm}(\xi^s_{;s}-1)
+2\alpha_2g^{sm}\xi^b_{;s}\big]F_{ab}F_{mn}.
\end{equation}
In our path towards the equations of motion of elastic matter
subjected to electrodynamic forces we are now confronted with two
options. We could attempt evaluating the Hamiltonian derivative of
${\bf L}_{(s.)}$  with respect to $g_{ik}$, then add the resulting
tensor density to the overall energy tensor density ${\bf
T}^{ik}$, of which we already know the inertial term from
(\ref{5.12}) , the elastic part from (\ref{5.11}), and the
``unstrained'' electromagnetic component (\ref{6.11}). The
vanishing divergence of ${\bf T}^{ik}$ would then provide the
equations of motion for the fields $\xi^i$ and $\rho$, once the
appropriate substitutions have been made, in keeping with the
definition (\ref{5.5}) of the four-velocity $u^i$. This program
meets however with a certain difficulty: it requires assessing the
metric content of $\xi^i_{;k}$ through extra assumptions of
arbitrary character.\par  An alternative way is however offered.
One can determine directly, without extra hypotheses, the
contribution to the generalized force density ${\bf f}_{i(s.)}$
stemming from electrostriction through the Euler-Lagrange
procedure:
\begin{equation}\label{6.18}
{\bf f}_{i(s.)}={\partial {\bf L}_{(s.)}\over\partial\xi^i}
-{\partial\over\partial x^k}\big({\partial {\bf
L}_{(s.)}\over\partial\xi^i_{~,k}}\big).
\end{equation}
If the metric $g_{ik}$ is everywhere given by equation
(\ref{3.1}), and the velocity is so small that $u^\rho$ can be
dealt with as a first order infinitesimal quantity, while $u^4$
differs from unity only for second order terms, the Lagrangian
density (\ref{6.17}) comes to read:
\begin{equation}\label{6.19}
L_{(s.)}={1\over2}\alpha_1F_{4\sigma}F^{4\sigma}\xi^\rho_{,\rho}
-\alpha_2F_{4\rho}F_{4\sigma}\xi^{\rho,\sigma},
\end{equation}
and the nonzero components of its Hamiltonian derivative with
respect to $\xi^i$ are:
\begin{eqnarray}\label{6.20}
{\bf f}_{\rho(s.)}={\delta L_{(s.)} \over {\delta \xi^\rho}}
=-{\partial \over {\partial x^\nu}} \bigg({\partial L_{(s.)} \over
{\partial \xi^\rho}_{,\nu}}\bigg)\nonumber\\
=-{1\over2}\alpha_1\big(F_{4\sigma}F^{4\sigma})_{,\rho}
-\alpha_2\big(F_{4\rho}F^{4\nu}\big)_{,\nu}.
\end{eqnarray}
In the case of fluid matter $\alpha_2$ is vanishing, and the
expression of the force density given by this equation agrees with
the one predicted long ago by Helmholtz with arguments about the
energy of an electrostatic system \cite{Helmholtz1881}, and
vindicated by experiments \cite{Goetz58}, \cite{
Zahn62} performed
much later.
\section{Conclusive remarks}
\label{sec:7}The results of the previous Sections can be availed
of in several ways. The full general relativistic treatment would
require a simultaneous solution of Einstein's equations, of
Maxwell's equations and of the equations ${\bf T}^{ik}_{~;k}=0$
fulfilled by the overall energy tensor, thereby determining in a
consistent way the metric $g_{ik}$, the four-potential
$\varphi_i$, the ``displacement'' four-vector $\xi^i$, and $\rho$.
This approach is presently {\it extra vires}, due to our ignorance
of the part of ${\bf T}^{ik}$ stemming from ${\bf L}_{(s.)}$, for
the reason mentioned in the previous Section. Achievements of
lesser consistency are instead at hand, like solving the equations
for the electromagnetic field and for the material field described
by $\xi^i$ and $\rho$ with a given background metric, or finding
the motion of elastic matter with a fixed metric, while the
electromagnetic field is evaluated as if electrostriction were
absent. Obviously enough, the calculations become trivial when the
metric is everywhere given by (\ref{3.1}), while the motion of
matter occurs with non relativistic speed. In the present paper we
have required that matter be isotropic when at rest and
unstrained, but this limitation was just chosen for providing a
simple example. The theory can be extended without effort to
cristalline matter exhibiting different symmetry properties, for
which reliable experimental data have started accumulating in
recent years.

\end{document}